\documentclass{article}

\usepackage[preprint,nonatbib]{neurips_2020}

\usepackage[utf8]{inputenc} 
\usepackage[T1]{fontenc}    
\usepackage{hyperref}       
\usepackage{url}            
\usepackage{booktabs}       
\usepackage{amsfonts}       
\usepackage{nicefrac}       
\usepackage{microtype}      
\usepackage{xcolor} 	

\usepackage{enumerate}
\usepackage{amssymb}
\usepackage[export]{adjustbox}
\usepackage{wrapfig}
\usepackage{longtable}
\usepackage{subcaption}
\usepackage{enumitem}

\newcommand{\eat}[1]{}
\usepackage[ruled,vlined]{algorithm2e}
\usepackage{amsfonts}
\usepackage{amsmath}
\usepackage{graphicx}
\usepackage{float}
\usepackage{listings}
\usepackage{amsthm}
\usepackage[square,numbers,compress]{natbib}

\definecolor{YellowOrange}{HTML}{F1A348}
\definecolor{MedGreen}{HTML}{008000}

\usepackage[backgroundcolor=white, disable]{todonotes}

\title{Firenze: Model Evaluation Using Weak Signals} 

\author{%
		Bhavna Soman, Ali Torkamani, Michael J. Morais, Jeffrey Bickford, Baris Coskun\\
		Amazon Web Services\\
		\texttt{\{bhsoman, alitor, moraismi, jbick, barisco\}@amazon.com} \\
}

\begin{document}
\maketitle

\raggedbottom

\begin{abstract}	
	Data labels in the security field are frequently noisy, limited, or biased towards a subset of the population. As a result, commonplace evaluation methods such as accuracy, precision and recall metrics, or analysis of performance curves computed from labeled datasets do not provide sufficient confidence in the real-world performance of a machine learning (ML) model. This has slowed the adoption of machine learning in the field. In the industry today, we rely on domain expertise and lengthy manual evaluation to build this confidence before shipping a new model for security applications. 
In this paper, we introduce Firenze, a novel framework for comparative evaluation of ML models' performance using domain expertise, encoded into scalable functions called {\it markers}. We show that markers computed and combined over select subsets of samples called {\it regions of interest} can provide a robust estimate of their real-world performances. Critically, we use statistical hypothesis testing to ensure that observed differences---and therefore conclusions emerging from our framework---are more prominent than that observable from the noise alone. Using simulations and two real-world datasets for malware and domain-name-service reputation detection, we illustrate our approach's effectiveness, limitations, and insights. Taken together, we propose Firenze as a resource for fast, interpretable, and collaborative model development and evaluation by mixed teams of researchers, domain experts, and business owners.

\end{abstract}

\section{Introduction}
\label{sec:intro}
In research areas like information security in which data abounds but domain-expert labels or annotations for those data are uniquely expensive \citep{JoyceEtal2021}, reliable evaluation of a machine learning model's performance is challenging \citep{paxsonpaper, zhu2020measuring, avLabels15}. When developing models for use in real-world production environments from such restrictive datasets, how can we determine whether a newly-developed model will {\it actually} perform better than existing methods when deployed? Problematically, ``better'' is a problem-specific and multifarious consideration, {\it e.g.} a classifier is better if it provides a higher true positive detection rate than existing methods without exceeding a maximum-tolerable false positive rate. This remains an active research problem and barrier to the productionization of machine learning models to solve real-world problems \citep{paxsonpaper}.

Machine learning for information security, {\it e.g.} for malware or network intrusion detection, are of growing interest in both academia and industry \cite{bilge2011exposure, antonakakis2010building, lison2017neural, ramzanreputation, malwareClassificationNNSaxe2015, malwareClassificationAdversarialRobust2018, malwareClassificationAnomaly}. But, to illustrate the challenge, consider a case study of training a malicious domain name classifier from a list of malicious and benign domain names obtained from third-party threat intelligence providers \cite{PublicBlocklist}. Such labeled datasets are often restrictive, non-representative subsamples of much larger populations, and cannot faithfully capture the underlying data distribution. If we computed canonical evaluative metrics like precision and recall on these data, we could arbitrarily mis-estimate our model's performance on the billions of unlabeled domain names it would observe each day in production. Indeed, empirically, it is not uncommon for a model developed in this way to yield unreasonably high numbers of false positives in production, thereby rendering itself useless despite achieving near-perfect precision and recall scores on labeled data. Worse still, any effortfully-collected labeled data carries a high risk of becoming obsolete, since adversaries frequently change tactics and move targets. Taken together, these pecularities can induce severe concept drift, label drift, {\it and} covariate shift \cite{malwareConceptDrift, pendlebury2019tesseract, barbero2022transcendent}.

As a result, reliable evaluation of ML models in information security has required extensive manual investigations by qualified experts with highly specialized training. In this paper, we present Firenze, a novel model evaluation framework to automate this investigative process by scalably operationalizing their domain expertise using weak signals, and using these to compare models' performances without ground-truth labels. Our goal is to accelerate the iterative development process of machine learning models, and empower a collaborative workflow between research scientists, domain experts, and business owners solving emergent problems in information security.

Firenze encodes domain expertise into a set of rules, called markers. These markers collectively and scalably represent benchmarks, heuristics, and/or knowledge that would be used by domain experts during manual investigations of the outcomes of an ML-based model. They are {\it weak signals}, insofar as they may not be correct for every individual datum, but provide reliable insights over populations of data and/or consensuses of multiple markers. We apply these markers to a [unlabeled] test dataset to make principled judgments of the performance of that model with respect to some existing methods. Specifically, we measure how much higher and lower it is able to rank datapoints associated with malicious and benign markers {\it resp.}, compared to those other methods; using statistical hypothesis tests on specific regions of the test dataset, that indicate when a proposed model is {better than}, {worse than}, or {not different than} existing methods. By construction, the results from each individual marker readily provide a semantic understanding of why, how, and on which data model improvements or deteriorations are occurring. As a result, Firenze provides a nuanced picture of the comparative model performance along with the overall judgment of which model is more performant.

In Section \ref{sec:related}, we review related work from semi-supervised, unsupervised, weakly supervised learning, and information security. In Section \ref{sec:method}, we describe Firenze in detail, and in Section \ref{sec:simulations}, we investigate the efficacy of our approach using simulated data. In Section \ref{sec:experiments-malware} and \ref{sec:experiments}, we present two case-study applications of our approach to real-world datasets. In Section \ref{sec:conclusion}, we discuss future directions and intersectional opportunities for our work.

\section{Related work}
\label{sec:related}
Evaluation of machine learning models in the security literature broadly uses canonical metrics like accuracy, precision, and recall {\it on labeled data} \cite{paxsonpaper}, but such approaches can be inaccurate or limited for data with partial labels, noisy labels, or no labels at all \citep{fedorchukclassifier, ratner2017data}. For example, a metaanalysis of Android malware classifiers by \citet{pendlebury2019tesseract} revealed that their accuracies were heavily biased over time and across groups of samples. In turn, there is a growing emphasis on using real-world and/or high-quality datasets \citep{shiravi2012toward, anderson2018ember} and seeking explainable, semantic understanding of model outcomes \citep{arp2014drebin}.


Naively, one could improve the quality of evaluative metrics like these by obtaining reliable ground truth labels. For example, image and text corpora for simple labeling tasks like object recognition or text sentiment analysis can be labeled through large-scale crowdsourcing campaigns \citep{crowdsourcing-book} using Mechanical Turk (or similar services) \citep{mTurk}. However, such efforts do not scale to specialized labeling tasks like those in information security, which require investigations by a select few domain experts with rigorous training and experience \citep{JoyceEtal2021}.

For some datasets, ``labels'' can be retrieved from aggregators like VirusTotal or various threat intelligence feeds, but implicit assumptions of their reliability have been questioned \citep{zhu2020measuring, li2019reading}. Even when multiple such labeling sources are available, how to combine them into a single label, {\it i.e.} by a hard threshold/criterion \citep{zhu2020measuring} or model-based pooling \citep{avLabels15}, can be highly consequential in the resulting label and model outcomes trained upon them \citep{zhu2020measuring}. 

The nascent field of weak supervision has emerged in response to this problem of intractable, costly, and/or imprecise data labeling \citep{ratner2019weak}. The Snorkel project \citep{Snorkel2017} introduced so-called labeling functions to generate training datasets based on weak domain expert signals, and has been adopted for real-world problems spanning banking and biotech to remove the human labeling problem \citep{snorkelcasestudies}. This work has only recently been applied to information security, specifically for Windows malware classification \citep{tully2019sifter}. 

Directly targeting model evaluation instead, \citet{Deng2021CVPR} proposed AutoEval to estimate the accuracy of a classifier on an unlabeled dataset by using feature statistics from the training set and synthetic datasets generated by applying transformations to the training set. \citet{czechTranslation} use density estimation to similar ends. Most recently, \citet{JoyceEtal2021} define {Approximate Ground Truth Refinements} to estimate bounds on precision and recall over partially-labeled data. All of the methods discussed thusfar augment the {\it training} process in some way; we propose Firenze as a black-box method that can directly evaluate an already-trained model without retraining. 


\section{Firenze: Model evaluation using weak supervision}
\label{sec:method}

\begin{figure*}[t]
	\centering
	\includegraphics[width=\textwidth]{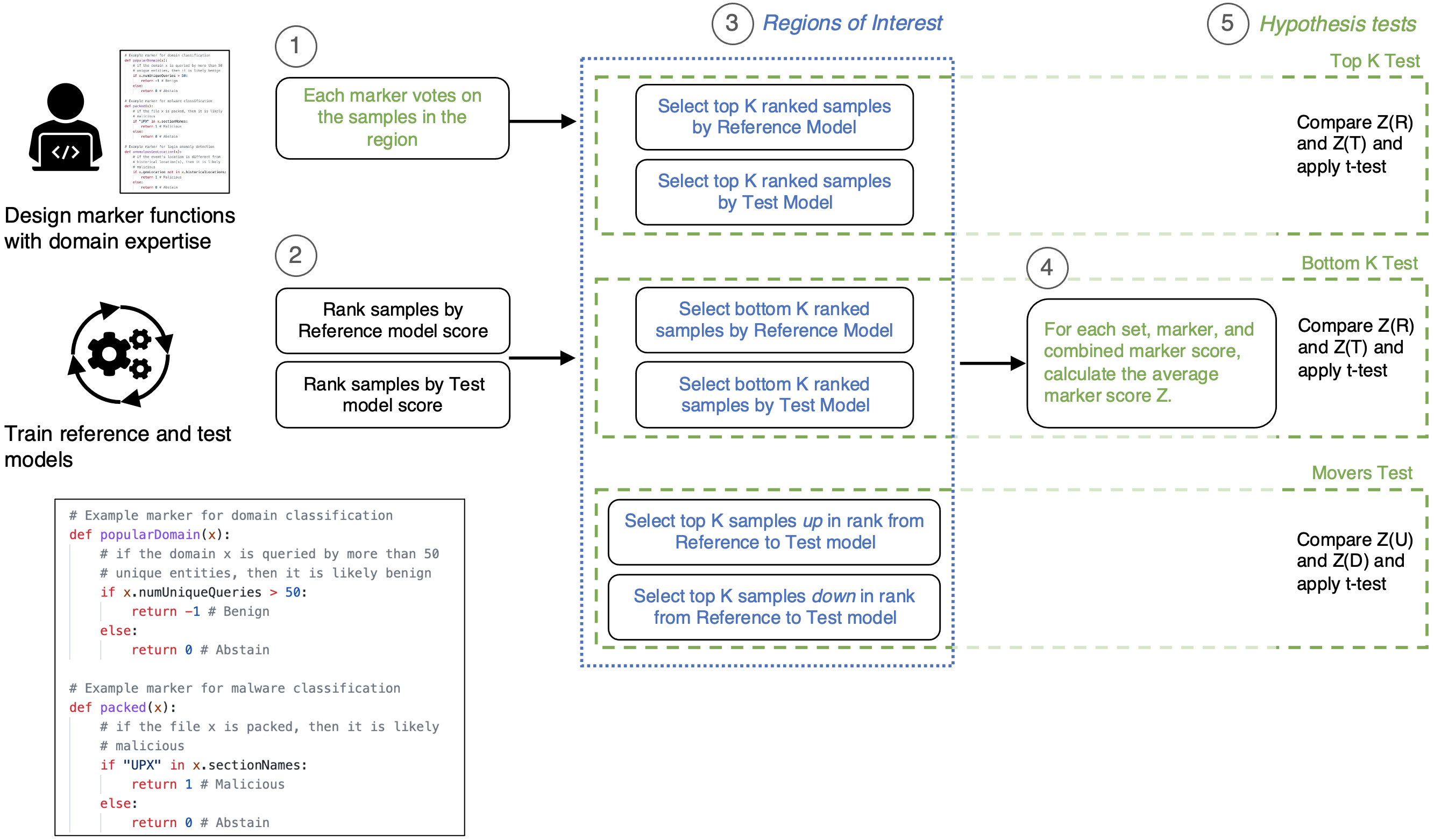}
	\caption{An overview of the Firenze system. (1) A domain expert defines the marker functions. (2) Create ranked lists of the samples by each model. (3) Assign samples to regions of interest. (4) Calculate the average marker score per set. (5) For the two sets in each region of interest, determine the better model by comparing the average marker scores using a two-sample unequal-variance t-test. {\it Inset:} Examples of marker functions for applications of ML in security}
	\label{fig:BlockDiagram}
\end{figure*}

Firenze is a framework for pairwise, comparative model evaluation utilizing weak signals for both supervised (e.g.~classification of malicious vs.~benign domain names) and unsupervised score-based models (e.g.~ anomaly detection). Firenze also attempts to bridge the semantic gap by using domain expert weak signals to describe semantically how a model is performing outside of ground-truth labels addressing the recurring concerns of ML models in information security \citep{paxsonpaper}. At a high level, our goal is to compare an existing model ({\it i.e.} one in production, {\bf Reference Model}) with a newly built model ({\bf Test Model}). Firenze features the following components, as summarized in Figure \ref{fig:BlockDiagram}:

These constituent parts evaluate and compare two models, which we denote Model $R$ (or \emph{Reference Model}) and Model $T$ (or \emph{Test Model}). These models need only share a common goal/task, {\it e.g.} classifying malware or domain names; they may differ in feature representations of their input data, model architectures, etc. Critically, Firenze performs its evaluations strictly on the output scores of these models. Such a black-box treatment of these models permits fast, easy incorporation into [existing] research pipelines and well-posed comparisons of diverse models.

\subsection{Marker design and combination}\label{section:Markers}
A marker is a weak signal that is associated with the maliciousness or benignity of a sample, instance, or event. The weak signal can come from diverse sources, patterns, heuristics and external knowledge bases that operationalize a security expert's intuition of whether a sample is malicious or not. These intuitions may not be correct for every individual case, but broadly hold true for the population. 
For example, malware analysts understand that not all packed files are malicious, but the fact that a file is packed is at least suspicious and increases the odds of it being malicious. Similarly, when analyzing a suspicious, recently-queried domain name, analysts may check how popular this domain has been over some time window. 


We define $M$ marker functions ${m_1,...,m_M}$ where $m_j(s)$ indicates the verdict of the $j^{th}$ marker (if any) observed for a sample $s$. 
Markers' verdicts span $m_j(s) \in \{-1, 0, 1\}$, where $-1$ indicates that the marker voted the sample $s$ to be benign, $1$ indicates malicious and $0$ indicates that the marker abstains. Allowing markers to abstain is important as the opposite of a security expert's intuition does not always indicate a vote for the opposite class. Revisiting our packed malware example above, {\it not} being packed is a poor weak signal of benignness---numerous malware aren't packed. 



By design, a single marker may not give a conclusive verdict for a sample's maliciousness or benignness; however, a combination of many such markers can provide a stronger overall verdict, and emulate how human experts build confidence and make inferences. To aggregate individual markers, we define the  \emph{combined marker score} as their majority vote, which itself can ``abstain'' with $0$ for ties. While this is a naive method, past research has shown that in use cases with low signal density (like ours) there is limited room for even an optimal weighting of the signals to diverge much from the majority vote \cite{Snorkel2017}. More sophisticated aggregation based on Dawid-Skene estimators \cite{DawidSkene} or generative models are planned for future work. 

We still expect the combined marker score of individual samples to be noisy; which is why we compare populations in our evaluation. Over the subsets/regions of samples considered below, we calculate the \emph{average marker score} of the samples in a given set, denoted $Z(R)$ and $Z(T)$ for models $R$ and $T$ {\it resp.}. Intuitively, if a set contains more samples that are likely malicious, its average marker score will be greater, and vice versa for fewer samples. 

\subsection{Region-based hypothesis testing} 
ML models in the security domain generally seek a robust separation of malicious and benign samples, but may only use a limited range of their operation. For example, a domain name reputation model may score millions of unique domain names per day, but only a small [fixed-size] subset of those will be sufficiently [confidently] benign to allowlist. 
Consequentially, which samples a model places in such {\it regions of interest} becomes instrumental its real-world performance. Samples for which the assigned ``region'' changes from one model to the next grant evaluative information about the comparative performance of the two models. Therefore, we propose to perform comparative evaluation on three regions of interest of size $K$: one each to explore the ``most malicious'' samples, ``most benign'' samples, and most differently-scored samples; other such regions may exist for use-cases not considered here. 

For each model, these regions are defined by their output {\it scores} $p=\text{Model}(s)$ assigned to input samples from a test dataset, which reflects some confidence or probability that each sample belongs to the malicious or benign class. These scores need not be comparable directly across models, {\it i.e.} SVM margin scores or class probabilities. Instead, we sort samples by their scores in each model, such that samples' {\it ranks} are comparable across models. Across a large set of [unlabeled] test data, we can associate each sample with ({\it i}) its rank score and ({\it ii}) its marker score. These two scores define Firenze's {\it tests}: 

\begin{itemize}[noitemsep,nolistsep,leftmargin=*]
\label{subsec:testDefinition}
	\item {\bf Top-$K$ Test}: We hypothesize that, within the top-$K$-ranked samples by model score for some $K$, the test model is {\it better} than the reference model if it assigns more {likely malicious} samples and fewer {likely benign} samples to this region. As defined by the marker scores above, this is tantamount to testing whether $Z(T)>Z(R)$, {\it i.e.} whether Model $T$ has a {\it higher} average marker score in this region. 
	\item {\bf Bottom-$K$ Test}: Conversely, we hypothesize that, within the bottom-$K$-ranked samples by model score for some $K$, the test model is {\it better} than the reference model if it assigns more {likely benign} samples and fewer {likely malicious} samples to this region. Likewise, we test whether $Z(T)<Z(R)$, {\it i.e.} whether Model $T$ has a {\it lower} average marker score in this region. 
	\item {\bf Movers Test}: We hypothesize that the test model is {\it better} than the reference model if it assigns {\it more malicious} (as defined by marker score) samples to higher ranks (as defined by model score), and {\it more benign} samples to lower ranks. Specifically, for some $K$, we use model scores to select the $K$ samples with largest increase in rank from Model $R$ to Model $T$---``up-movers''---and the $K$ samples with largest decrease---``down-movers''. Then, we test $Z(U)>Z(D)$, {\it i.e.} whether the average marker score of up-movers is higher than that of down-movers. 
\end{itemize}

For each of the Top-K, Bottom-K, and Movers Tests, we compare the average marker scores of the samples placed in each region by Models R and T using a two-sample t-test with unequal variance at level 0.05, called Welch's t-test \cite{welchTest}. This permits us to observe and interpret differences in $Z(R)$ and $Z(T)$ sensitive to variability in these estimates, only if we can exclude the uninformative statistical possibility that the observed differences arose by random chance between equally performant models (probability $p \leq 0.05$). In other words, if the difference in average marker scores is larger than that which could be observed by an overwhelming fraction of chance outcomes, then we conclude the two models are performing statistically differently over that region. 
In practice, we run these statistical tests as {\it two-sided tests} of whether $Z(T)\neq Z(R)$ and $Z(U)\neq Z(D)$, rather than a one-sided test of whether one is greater-than or less-than the other; in doing so, we can also identify when the test model is {\it worse} than the reference model, by the same hypotheses above. 

\section{Assessing Firenze using simulated data}
\label{sec:simulations}

To demonstrate Firenze on data and models with known ground-truth labels, we developed an extensive simulated environment that parametrizes and partitions relevant sources of noise endemic to a model training-and-testing pipeline. Details of the generative process are given in Appendix \ref{app:simulations}. Our goal is to define qualitative conditions---in terms of the simulations' parameters---under which Firenze can identify the better model with the proposed region-based hypothesis tests. This is a function of markers' accuracies and coverages, and their relationship to other parameters of this simulated environment, like the accuracy and coverage of the noisy, incomplete training labels with respect to some ground truth. In specific contrast to real-world datasets, only in simulations can we disambiguate between differences in objective, unobserved ground-truth labels and subjective, observed training labels, and how these propagate to model performance and our evaluation of it. 

Key features of this simulation are {\it (i)} generation of ground-truth labels as well as noisy generation of training labels and weak signals (markers) of arbitrary accuracy and coverage (with respect to ground-truth), and {\it (ii)} model score generation with arbitrary performances with respect to either of the labelsets. With these features, we explore the requirements of a single marker, knowing that these results provide a lower bound on any other use-case.


\subsection{Experiments}


The parameters for our simulated environment are the positive class prevalence $\pi$, reference and test model performances on ground-truth and training labels $P_\text{true}^R$, $P_\text{train}^R$,$P_\text{true}^T$, and $P_\text{train}^T$, the number of samples $N$, the region size $K$, and the accuracies and coverages of the marker $\alpha$ and $\beta$ {\it resp.} and the training labels $\bar{\alpha}$ and $\bar{\beta}$ {\it resp.}. Their default values, unless specified otherwise, are $P_\text{train}^T=0.97$, $P_\text{train}^R=0.98$, $P_\text{true}^T=0.95$, $P_\text{true}^R=0.90$, $\pi=0.5$, $\overline{\alpha}=0.95$, $\overline{\beta}=0.10$, $K=10000$, and $N=1000000$. The choices focus our experiments on the most nefarious case of model evaluation: the training performances are fixed such that $P_\text{train}^T < P_\text{train}^R$, the opposite of the true difference on ground-truth labels where $P_\text{true}^T > P_\text{true}^R$.\todo{{\it DONE, we ``simulate'' training, see previous section} why do we care about training labels here? We don't train models. We just evaluate.}

\begin{figure*}[t]
	\centering
	\includegraphics[width=0.475\linewidth]{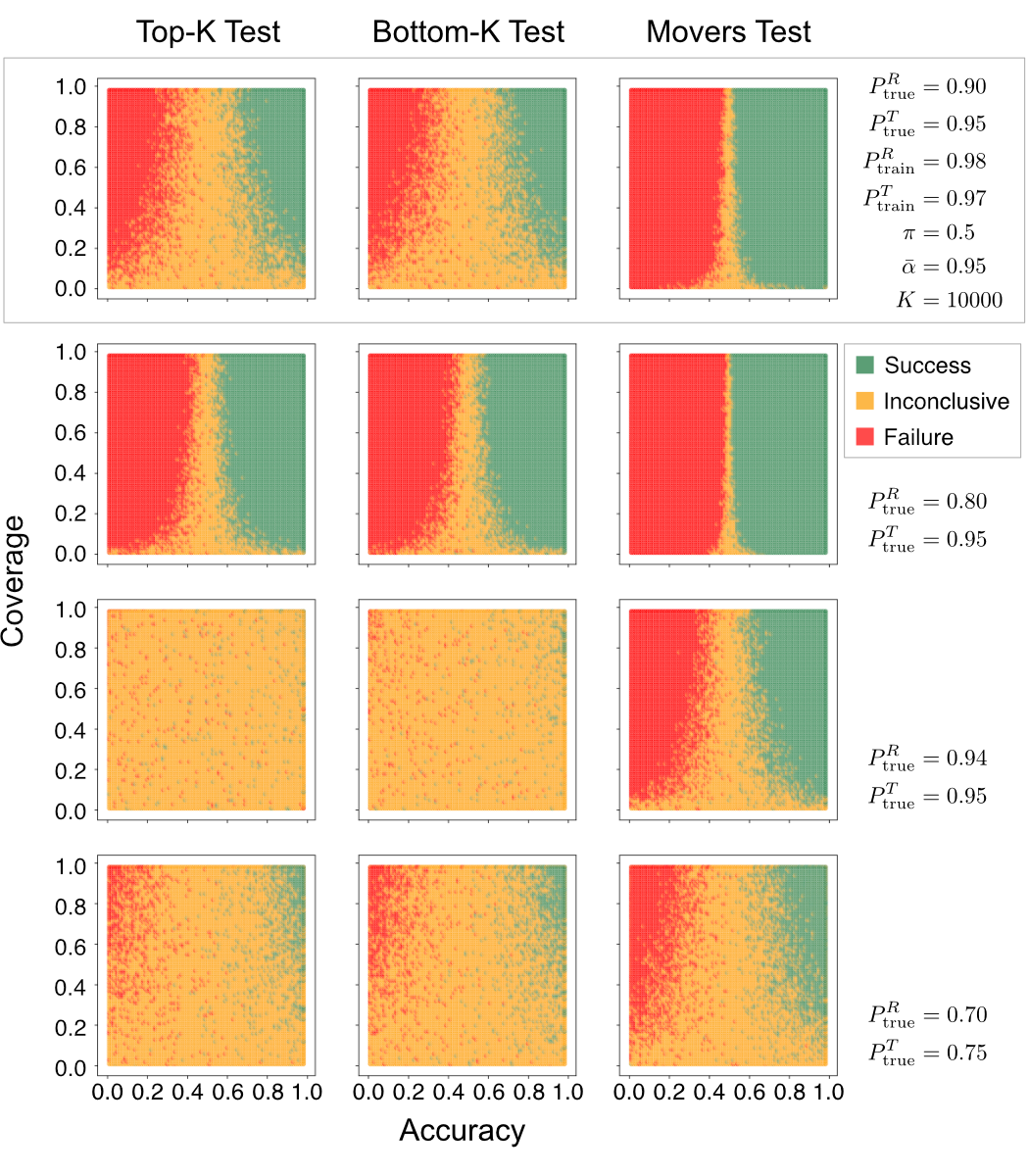}
	\includegraphics[width=0.475\linewidth]{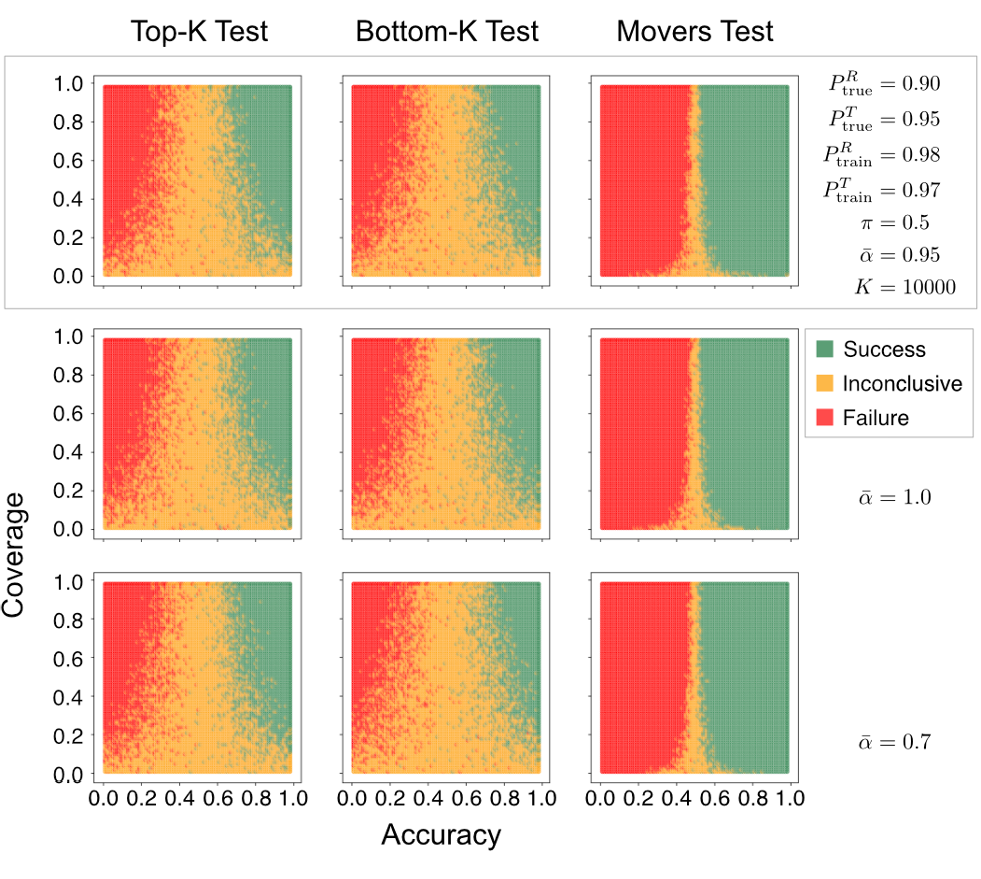}
	\caption{Varying difference in model performances $P^T_\text{true}$ and $P^R_\text{true}$ (left) and feed accuracies $\bar{\alpha}$ (right). Using the default simulation parameters as a guide (top, in boxes), in all panels we observe test Success (green) as marker accuracy increases $\alpha>0.5$ and Failure (red) as marker accuracy decreases $\alpha<0.5$ ({\it x-axis}). The interior region is Inconclusive (yellow), and that region widens---the test becomes less sensitive---as marker coverage decreases ({\it y-axis}). {\it Left,} all tests become more (less) sensitive as the true difference in performance becomes larger (smaller). {\it Right,} Test sensitivity does not depend on the accuracy of the training labels.}
	\label{fig:sim_P_alpha}
\end{figure*}

\begin{figure*}
	\centering
	\includegraphics[width=0.475\linewidth]{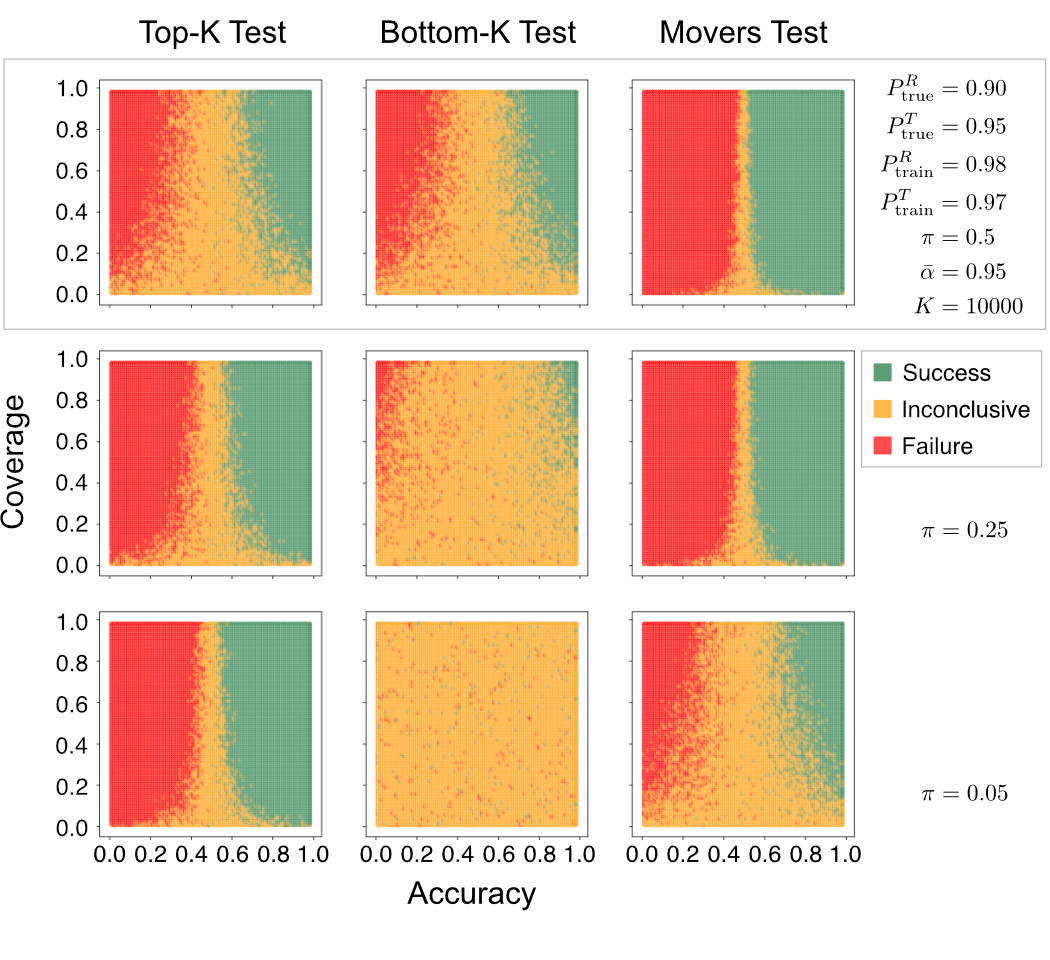}
	\includegraphics[width=0.475\linewidth]{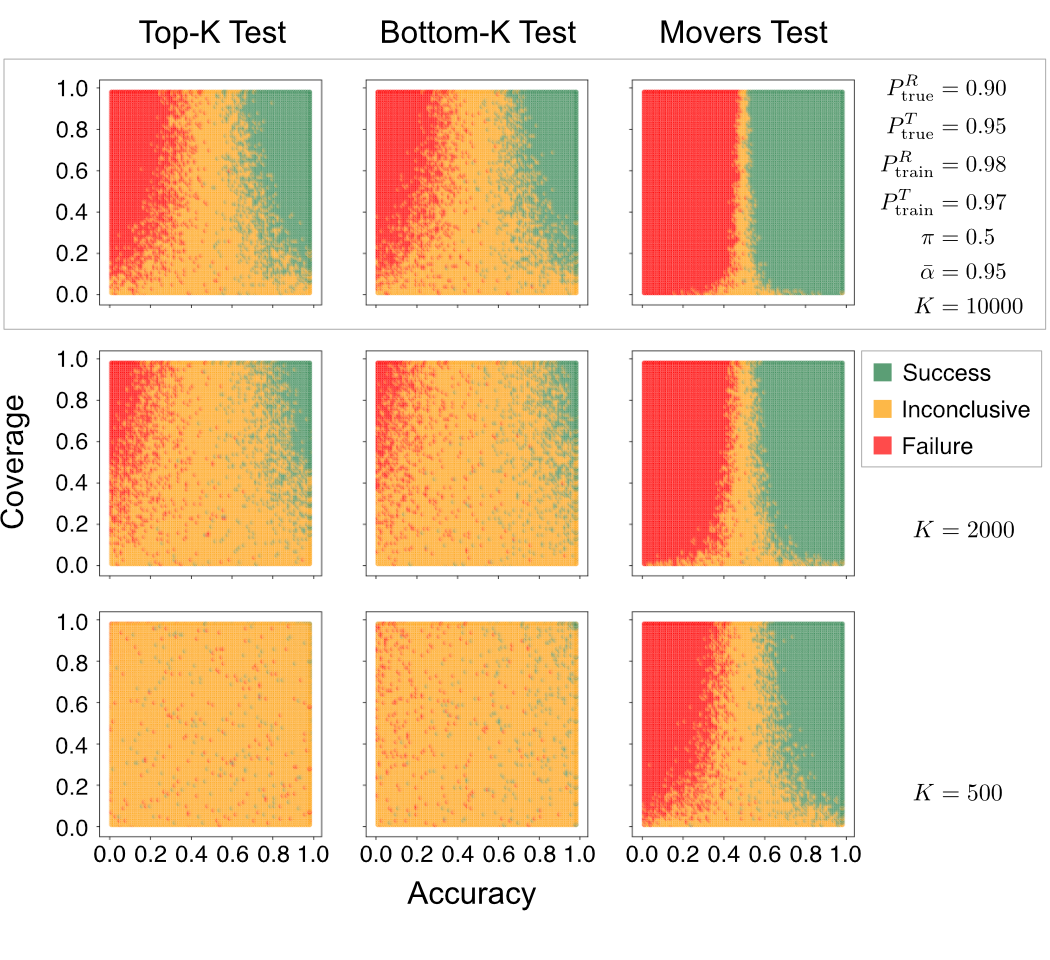}
	\caption{Varying class prevalence $\pi$ (left) and ROI set size $K$ (right; {\it cf.} Fig. \ref{fig:sim_P_alpha} for how to interpret the panels). Both parameters have asymmetric effects on the regions. {\it Left,} as positive class prevalence decreases, the Bottom-K and Movers Tests lose sensitivity, while the Top-K gains sensitivity. {\it Right,} as region size decreases, all tests uniformly lose sensitivity.}
	\label{fig:sim_pi_K}
\end{figure*}

Holding all other parameters to their default values, we explore the role of the ground-truth model performances $P$ (Fig. \ref{fig:sim_P_alpha}, {\it left}), training label accuracy $\bar{\alpha}$ (Fig. \ref{fig:sim_P_alpha}, {\it right}), positive class prevalence $\pi$ (Fig. \ref{fig:sim_pi_K}, {\it left}), and region size $K$ (Fig. \ref{fig:sim_pi_K}, {\it right}).\todo{{\it ATTEMPTED}: Overall experiment design is not coming out clear. We need to make another pass and elaborate a little bit more.} In each experiment, for a given parameter configuration, we simulate this process for each of $N$ samples, generating true labels, then training labels and model scores for reference ($R$) and test ($T$) models, and finally markers. Using the model scores and markers, we apply the Firenze framework, and observe the outcomes of the three significance tests to identify the model with higher ground-truth performance. 

Given our goal---to study the minimal requirements of markers---we repeat this simulation on a fine tiling of marker accuracies $\alpha\in(0, 1)$ and coverages $\beta\in(0, 1)$, and plot the result of each at the coordinate $(\alpha, \beta)$ in each figure panel to follow. The resulting visualization shows the success, failure, and inconclusive regimes of the Firenze tests, as a function of the marker's parameters. In each figure, each column of panels reflects a certain region/test (Top-K, Bottom-K, Movers), and row of panels reflects a certain parameter configuration (annotated accordingly). 


{\bf Ground-truth model performance.} For fixed model performances on training data, we varied the model performances on ground-truth data $P_\text{true}^T$ and $P_\text{true}^R$ (Fig. \ref{fig:sim_P_alpha}, {\it left}). Relative to the default model, a larger difference ($P_\text{true}^T=0.95$ vs. $P_\text{true}^R=0.80$) with low generalization error ($P_\text{true}^T=0.95$ vs. $P_\text{train}^R=0.98$) increases sensitivity of all tests. A small difference ($P_\text{true}^T=0.95$ vs. $P_\text{true}^R=0.94$) decreases sensitivity of all tests, and to a lesser degree of \eat{except} the Movers test\eat{, suggesting the Movers Test is identifying the ~$1\%$ of samples with differing predictions}. The default difference with high generalization error ($P_\text{true}^T=0.75$ vs. $P_\text{true}^R=0.70$) strongly decreases sensitivity of all tests. 


{\bf Training label accuracy.} We then varied the reliability of training labels, which in turn varies the generalization errors of our two models (Fig. \ref{fig:sim_P_alpha}, {\it right}). Because markers are independent of the noise level in the training labels, this does not impact test sensitivity for any test nor any accuracy level. We emphasize that the lack of dependence on training label accuracy underpins the power of these tests.

{\bf Positive class prevalence and region size.} Finally, we varied class prevalences $\pi$ and region sizes $K$ to explore dependence on the sample data balance and size (Fig. \ref{fig:sim_pi_K}). As the positive (malicious) class becomes more rare in the dataset, the Top-K test remains sensitive, as the top-K samples can still contain adequate sample counts for both positive and negative classes; the Bottom-K and Movers Tests both lose sensitivity for the converse reason, as their samples will be overwhelmingly negative. As the region size $K$ decreases (reducing training and evaluation set sizes equally), all tests lose sensitivity, though least so for the Movers Test.

\subsection{Qualitative conditions for successful tests}

Varying the parameters of this simulated environment modulates the sensitivity of the tests in the Firenze framework. Importantly, none of these regimes bias the tests, therefore as long as the markers have accuracy $\alpha>0.5$, Firenze can yield {\it at worst} an inconclusive result, at best a success. Qualitatively, we observe that, when evaluating highly-performant, incrementally-different models (all $P>0.9$), a single marker with accuracy $\alpha>0.7$ and coverage $\beta>0.5$ can successfully identify the better model with reasonable probability. \eat{From Fig. \ref{fig:majority_voting_fixed_alpha}, we further observe that the same results could be achieved with more markers of lower accuracy, e.g. $7$ markers with $\alpha>0.6$, or better results could be achieved with more markers of the same accuracy.}

The other parameters we varied suggests a loose ``operating regime'' for evaluation with Firenze. Within user control, large(r) region-of-interest sizes $K$ yield more sensitive tests. Outside user control, low positive-class sample size $\pi$, significant generalization errors $P_\text{true}\ll P_\text{train}$, and/or small differences in ground-truth performance $P_\text{true}\approx P_\text{train}$ yield less sensitive tests, especially for Top- and Bottom-K Tests. Taken together with the higher sensitivity of the Movers Test throughout, these observations suggest that regions-of-interest yield successful tests when they have a heterogeneity of labels, i.e. a propensity for non-zero differences in marker score to emerge. We are optimistic that future work can affirm these relationships and insights analytically and provide a broader theory of evaluative weak signals.

\section{Evaluating malware detection models using Firenze}
\label{sec:experiments-malware}
To illustrate how Firenze can be used in practice, we share a
first case study, a replicable proof-of-concept comparing two models for ML-based malware detection which use the EMBER open-source malware dataset \citep{anderson2018ember}. The EMBER dataset is a curated set of malicious and benign Windows PE files for {\it static analysis}. The default feature representation of these data spans features from file headers, section information, file imports and exports, directory information, and byte entropy statistics.

To construct an ecologically valid case study, we sort the EMBER data by the date/time at which each sample was first observed, to train our reference and test models on ``past'' data (pre-December 2017), perform preliminary tests on ``present'' data (December 2017), and evaluate with Firenze on ``future'' data (2018) \citep{pendlebury2019tesseract}. For this purpose, we specifically use the unlabeled samples from the 2018 period. The {\it reference model} is a neural network classifier with the same architecture of Erdemir et al. in their experiments with the EMBER dataset \citep{erdemir2021adversarial}). The {\it test model} is a gradient-boosted decision tree with the same hyperparameters of Anderson et al. in the original EMBER paper \citep{anderson2018ember}. 

On ``present'' data, performances of the reference and test models appear highly comparable ($\text{AUC}_R=0.9981$ versus $\text{AUC}_T=0.9984$), but on future data, model performances are known to degrade---and will our estimates of them---as covariate shift, concept drift, and/or label drift mount \cite{malwareConceptDrift, pendlebury2019tesseract, barbero2022transcendent}. Using Firenze on the unlabeled dataset, we investigate to what extent the change in model architecture improves performance by ({\it i}) increasing true malicious file identifications (true positives) by the model, without increasing false positives and ({\it ii}) improving identification of benign files without increasing false negatives.

\subsection{Markers for malware detection}

We designed five markers to evaluate these models; like above, we outline them here, and discuss details in Appendix \ref{app:ember:markerdetails}. Recall that a verdict of $1$ is malicious, $-1$ is benign, and $0$ is null/abstain:
\begin{itemize}[noitemsep,nolistsep,leftmargin=*]
\item {\bf Suspicious Section Properties}: If the sample contains more than one executable or any writable-and-executable section, then $1$, else $0$
\item {\bf Unusual Number of Imported Funtions}: If the same contains fewer than 25 imports---less than the usual packed sample---then $1$, else $0$
\item {\bf Nonsensical Section Names}: If the sample contains a nonsensical section name, as determined by \texttt{nostril} \citep{Hucka2018}, then $1$, else $0$
\item {\bf Import of suspicious functions}: If the sample imports functions and libraries with commonly malicious functionality (see Appendix \ref{app:ember:importfunctions} for function details) , then $1$, else $0$
\item {\bf Signed}: If the sample is signed by a trusted source, then $-1$, else $0$
\end{itemize}
Consider the second marker, unusual number of imports, and how it reflects  our definition of markers as {\it weak signals}. Though very few imports---common for packed/obfuscated samples---is a good signal of suspiciousness, numerous imports is not a signal of legitimacy by negation. Likewise, many malicious samples aren't packed, and could contain any number of imports.

\subsection{Region-based testing with $K$ = 50,000 samples}


Region-based hypothesis testing with Firenze is mechanistically amenable to malware detection. Suppose the models' predictions are triaged by a security operations team with a limited bandwidth of manual investigations. If a detection count of $K$ = 50k samples filled that bandwidth approximately, it would be sensible to evaluate model performance only over that region within which impactful security decisions are made. We apply Firenze to evaluate our two malware detectors on regions of 50k samples, {\it e.g.} for a blocklist (malicious verdicts, Top-$K$), allowlist (benign verdicts, Bottom-$K$), or investigative list (Movers). 

The reference and test models are pretrained, and we report the outcomes of Firenze's region-based hypothesis tests in Table \ref{tab:EMBERsummary} below. Each table reports their combined marker scores $Z(\cdot)$ ({\it abbrev.} CMS) on each region and the $p$-value of the t-test that tests each hypothesis by which the test model would be better than the reference model (or the reference better than the test; see Section \ref{subsec:testDefinition}). We summarize the outcome of each test with an {\color{MedGreen}S} (Success) to show "test model out-performs reference model" ($p\le$0.05), an {\color{red}F} (Failure) to show "reference model out-performs test model" ($p\le$0.05 for the opposite outcome), and a {\color{YellowOrange}U} (Undetermined) to show an inconclusive outcome ($p>$0.05). 

We see that all of the Top-K, Bottom-K, and Movers tests succeed, {\it i.e.} the test model is uniformly better at scoring malicious and benign samples, as well as moving malicious/benign samples to higher/lower ranks. These results are more granular, interpretable, and therefore trustworthy than miniscule differences in AUC on labeled data, such that a security expert would recommend the test model over the reference.

The EMBER dataset presents two explicit opportunities to verify conclusions drawn from Firenze's tests. First, because the dataset also contains ~800k labeled samples from the 2018 period used for evaluation, we can verify with classical metrics that the test model is, indeed, out-performing the reference model ($\text{AUC}_R=0.9166$ versus $\text{AUC}_T=0.9371$), though both show degredation of performance over time. Second, because we could manually retrieve VirusTotal reports and labels (0 verdicts $\Rightarrow$ benign, $\ge$40 verdicts $\Rightarrow$ malicious \citep{anderson2018ember}) on these once-unlabeled samples now---four years later---we can verify our conclusions once more ($\text{Accuracy}_R=0.90$ versus $\text{Accuracy}_T=0.94$).

\begin{table}[H]
	\centering
	\resizebox{0.65\textwidth}{!}{%
		\begin{tabular}{l|r|r|r|c}
			\textbf{Test} & \textbf{Avg CMS}  & \textbf{Avg CMS}  & \textbf{p-value} & \textbf{Result} \\
			& \textbf{Reference Model}  & \textbf{Test Model}  &  &  \\
			\hline
			TopK Test, 50k & 0.11456 & 0.68445 & <10\textsuperscript{-16} & \color{MedGreen}S \\
			BottomK Test, 50k &	0.09788	& -0.16862 & <10\textsuperscript{-16} & \color{MedGreen}S \vspace{0.1in}\\
			\textbf{Test} & \textbf{Avg CMS}  & \textbf{Avg CMS}  & \textbf{p-value} & \textbf{Result} \\
			& \textbf{Up-Movers}  & \textbf{Down-Movers}  &  &  \\
			\hline
			Movers Test, 50k & 0.42884 & 0.00868	& <10\textsuperscript{-16} &	\color{MedGreen}S
		\end{tabular}

	}
\vspace{0.1in}\caption{Outcomes of Firenze's evaluative comparison of reference and test malware detection models with the Top-K, Bottom-K, and Movers tests for $K$ = 50k \label{tab:EMBERsummary}}
\end{table}

\section{Evaluating domain name reputation models using Firenze}
\label{sec:experiments}
We follow up with a second case study from a mature real-world use-case comparing two models for domain name reputation, which use fully anonymized passive DNS data obtained from a large cloud service provider. The exact details of these models are not the focus of this paper, but can be assumed similar to previous related work in this space \cite{bilge2011exposure, antonakakis2010building, lison2017neural, ramzanreputation}.

These domain name reputation models are used to identify malicious domains for threat detection as well as benign domains for false positive mitigation. The {\it reference model} is an already-in-use production version of the model. The {\it test model} is a proposed update to the model which adds additional features. Both models would score as many as one billion domains per day, but are only trained on a few million domains with known labels. This large discrepancy makes model improvements difficult to evaluate, since precision and recall across model versions compared to labels stays relatively stable (here, the test model scores slightly better; area under the ROC curve $\text{AUC}_R=0.98387$ versus $\text{AUC}_T=0.98527$). Using Firenze on all domains, we investigate to what extent the new feature addition achieved our goals to ({\it i}) increase true malicious domain identifications (true positives) by the model without increasing false positives and ({\it ii}) improve identification of benign domains without increasing false negatives.

%


We designed seven markers to evaluate these models; we outline them here, and discuss the relevant background information and domain expertise that motivated them in Appendix \ref{app:mithra:markers}. 
\begin{itemize}[noitemsep,nolistsep,leftmargin=*]
    \item {\bf Abused Domain}: If the domain is associated with a curated list of known-abused domains, then $1$, else $0$
    \item {\bf Sinkholed Domain}: If the domain is associated with a curated list of known-sinkhole IP addresses, then $1$, else $0$
    \item {\bf Honeypot Domain}: If the domain appears in in-house honeypot logs, then $1$, else $0$ 
    \item {\bf Domain Popularity}: If the domain is considered popular based on query counts, then $-1$, else $0$    \item {\bf Number of IPs}: If the domain maps to more than 50 unique IP addresses, then $-1$, else $0$
    \item {\bf Number of TTLs}: If the domain appears with more than 500 TTLs (Time to Live), then $-1$, else $0$
    \item {\bf Known Future Label}: If the domain is labeled malicious in the future labels, then $1$, if it is labeled benign, then $-1$, else $0$
\end{itemize}

Region-based hypothesis testing with Firenze is mechanistically amenable to the domain-name reputation problem as well. Analogously, suppose we applied Firenze to evaluate our two domain-name reputation models on regions with $K$ = 10k {\it or} 100k samples. The reference and test models are pretrained, and we report the outcomes of Firenze's region-based hypothesis tests In Table \ref{tab:DNSsummary} below. 

In Table \ref{tab:DNSsummary}, we first see that both Top-K tests fail, {\it i.e.} the test model is worse than the reference model at scoring malicious domains, but both Bottom-K tests succeed, {\it i.e.} it is better at scoring benign domains. The Movers test fails for 10k and is inconclusive for 100k, {\it i.e.} the test model does not move malicious samples to higher ranks, and benign samples to lower ranks. We explicitly note that ``Success'' and ``Failure'', as noted in the tables, qualifies whether the test model successfully outperforms the reference model. Overall, we conclude that {\it we failed to develop a better model, but succeeded in identifying it so with Firenze.}

\begin{table}[b]
	\centering
	\resizebox{0.65\textwidth}{!}{%
		\begin{tabular}{l|r|r|r|c} 
			\textbf{Test} & \textbf{Avg CMS}  & \textbf{Avg CMS}  & \textbf{p-value} & \textbf{Result} \\
			& \textbf{Reference Model}  & \textbf{Test Model}  &  &  \\
			\hline
			TopK Test, 10k & 0.617138 & 0.516348 & <10\textsuperscript{-16} & \color{red}F \\
			TopK Test, 100k & 0.570214 & 0.405806 & <10\textsuperscript{-16} & \color{red}F \\
			BottomK Test, 10k &	-0.5795	& -0.9835 &	<10\textsuperscript{-16} & \color{MedGreen}S \\\vspace{0.1in}
			BottomK Test, 100k &	-0.54655 & -0.67804	& <10\textsuperscript{-16} & \color{MedGreen}S \\
			\textbf{Test} & \textbf{Avg CMS}  & \textbf{Avg CMS}  & \textbf{p-value} & \textbf{Result} \\
			& \textbf{Up-Movers}  & \textbf{Down-Movers}  &  &  \\
			\hline
			Movers Test, 10k & 0.0026 & 0.0074	& 0.011 &	\color{red}F \\
			Movers Test, 100k & 0.00036 & 0.00016	& 0.296 &	\color{YellowOrange}U \\			
		\end{tabular}
	}	
\vspace{0.1in}\caption{Outcomes of Firenze's evaluative comparison of reference and test domain name reputation models with the Top-K, Bottom-K, and Movers tests for $K$ = 10k and 100k \label{tab:DNSsummary}}
\end{table}
%
%

Using traditional metrics over labeled data like AUC above, we observed that the test model was doing marginally better. But, with Firenze, we reveal a more nuanced picture of better benign detection and worse malicious detection, which reflects the plausible situation in which one model is not \emph{uniformly better} than another; instead, they each have regimes in which they perform better or worse. The granularity of {\it these} insights are what a security expert would need to recommend that a business owner not ship the new model, citing likely false negatives for a customer. Defining markers and regions helps identify these aspects of performance, automate their evaluation with the robustness of statistical tests, and make confident business decisions based on the outcomes.

\section{Discussion and conclusion}
\label{sec:conclusion}
With this paper, we introduced Firenze as a modular, extensible framework for {\it post-hoc} comparative model evaluation, that constitutes a novel approach to the problem of learning from data with noisy, unreliable, or absent labels (see Section \ref{sec:intro}). We showed how Firenze is driven by markers, weak signals encoding domain expertise, and how we can leverage their aggregate information over subsets of samples with statistical significance tests to compare the performance of two models without requiring ground-truth labels (Section \ref{sec:method}). We demonstrated its efficacy on simulated data (Section \ref{sec:simulations}), as well as on two real-world case studies from malware detection and domain-name reputation, which illustrate how a user should construct and interpret markers and regions (Sections \ref{sec:experiments} and \ref{sec:experiments-malware}). 

The framework also allows for flexibility in defining {\it both} markers and regions of interest to specialize performance [improvements] users want to measure. Once these are implemented for a given use-case, they can be seamlessly reused across arbitrary model refinements and changes---small hyperparameter adjustments or even complete architectural overhauls. 
In all cases, the outcomes from Firenze are explainable. Since comparative performance can be viewed for each marker and test, this can be used to provide feedback for targeted model refinements.


This said, Firenze does not remove the need to acquire labels; high-quality labeled datasets remain the premier means to develop effective models. Firenze gives {\it comparative} insights into model performance, and cannot infer the absolute performance differences that are achievable with fully labeled data.
Moreover, these insights hinges on the quality of markers designed by domain experts. A reasonable ground truth dataset will help the expert ensure that the markers meet or exceed the quality conditions we lay down in sections \ref{sec:method} and \ref{sec:simulations}. We suggest that effective applications of ML in the security domain require both: datasets with a high-quality [sub]set of labels for model training, and improved evaluation methods (like Firenze) to estimate improvement in performance on real-world data, much of which is unlabeled.


We are optimistic that Firenze can enable more holistic machine learning model development for research problems in information security 
by creating opportunities for {\it direct} participation by security researchers and business owners, as well as the usual ML scientists. A security researcher may design markers, regions, and/or tests like those proposed in section \ref{subsec:testDefinition} to curate the aspects over which one model may be performing better than another, adaptively with their [evolving] domain expertise. The business owner can survey customers for desired improvements to product or model performance, which can motivate additional markers, and so on.

\bibliographystyle{unsrtnat}
\bibliography{main}

\begin{thebibliography}{35}
\providecommand{\natexlab}[1]{#1}
\providecommand{\url}[1]{\texttt{#1}}
\expandafter\ifx\csname urlstyle\endcsname\relax
  \providecommand{\doi}[1]{doi: #1}\else
  \providecommand{\doi}{doi: \begingroup \urlstyle{rm}\Url}\fi

\bibitem[Joyce et~al.(2021)Joyce, Raff, and Nicholas]{JoyceEtal2021}
Robert~J. Joyce, Edward Raff, and Charles Nicholas.
\newblock A framework for cluster and classifier evaluation in the absence of
  reference labels.
\newblock \emph{Proceedings of the 14th ACM Workshop on Artificial Intelligence
  and Security}, Nov 2021.
\newblock \doi{10.1145/3474369.3486867}.
\newblock URL \url{http://dx.doi.org/10.1145/3474369.3486867}.

\bibitem[Sommer and Paxson(2010)]{paxsonpaper}
Robin Sommer and Vern Paxson.
\newblock Outside the closed world: On using machine learning for network
  intrusion detection.
\newblock In \emph{2010 IEEE Symposium on Security and Privacy}, 2010.

\bibitem[Zhu et~al.(2020)Zhu, Shi, Yang, Qin, Zhang, Song, and
  Wang]{zhu2020measuring}
Shuofei Zhu, Jianjun Shi, Limin Yang, Boqin Qin, Ziyi Zhang, Linhai Song, and
  Gang Wang.
\newblock Measuring and modeling the label dynamics of online anti-malware
  engines.
\newblock In \emph{29th USENIX Security Symposium (USENIX Security 20)}, pages
  2361--2378, 2020.

\bibitem[Kantchelian et~al.(2015)Kantchelian, Tschantz, Afroz, Miller, Shankar,
  Bachwani, Joseph, and Tygar]{avLabels15}
Alex Kantchelian, Michael~Carl Tschantz, Sadia Afroz, Brad Miller, Vaishaal
  Shankar, Rekha Bachwani, Anthony~D. Joseph, and J.~D. Tygar.
\newblock Better malware ground truth: Techniques for weighting anti-virus
  vendor labels.
\newblock In \emph{Proceedings of the 8th ACM Workshop on Artificial
  Intelligence and Security}, AISec '15, page 45–56, New York, NY, USA, 2015.
  Association for Computing Machinery.
\newblock ISBN 9781450338264.
\newblock \doi{10.1145/2808769.2808780}.
\newblock URL \url{https://doi.org/10.1145/2808769.2808780}.

\bibitem[Bilge et~al.(2011)Bilge, Kirda, Kruegel, and
  Balduzzi]{bilge2011exposure}
Leyla Bilge, Engin Kirda, Christopher Kruegel, and Marco Balduzzi.
\newblock Exposure: Finding malicious domains using passive dns analysis.
\newblock In \emph{Ndss}, pages 1--17, 2011.

\bibitem[Antonakakis et~al.(2010)Antonakakis, Perdisci, Dagon, Lee, and
  Feamster]{antonakakis2010building}
Manos Antonakakis, Roberto Perdisci, David Dagon, Wenke Lee, and Nick Feamster.
\newblock Building a dynamic reputation system for dns.
\newblock In \emph{USENIX security symposium}, pages 273--290, 2010.

\bibitem[Lison and Mavroeidis(2017)]{lison2017neural}
Pierre Lison and Vasileios Mavroeidis.
\newblock Neural reputation models learned from passive dns data.
\newblock In \emph{2017 IEEE International Conference on Big Data (Big Data)},
  pages 3662--3671. IEEE, 2017.

\bibitem[Ramzan et~al.()Ramzan, Seshadri, and Nachenberg]{ramzanreputation}
Zulfikar Ramzan, Vijay Seshadri, and Carey Nachenberg.
\newblock Reputation-based security.
\newblock https://docs.broadcom.com/doc/reputation-based-security-en.

\bibitem[Saxe and Berlin(2015)]{malwareClassificationNNSaxe2015}
Joshua Saxe and Konstantin Berlin.
\newblock Deep neural network based malware detection using two dimensional
  binary program features.
\newblock In \emph{2015 10th International Conference on Malicious and Unwanted
  Software (MALWARE)}, pages 11--20, 2015.
\newblock \doi{10.1109/MALWARE.2015.7413680}.

\bibitem[\'{I}ncer Romeo et~al.(2018)\'{I}ncer Romeo, Theodorides, Afroz, and
  Wagner]{malwareClassificationAdversarialRobust2018}
\'{I}\~{n}igo \'{I}ncer Romeo, Michael Theodorides, Sadia Afroz, and David
  Wagner.
\newblock Adversarially robust malware detection using monotonic
  classification.
\newblock In \emph{Proceedings of the Fourth ACM International Workshop on
  Security and Privacy Analytics}, IWSPA '18, page 54–63, New York, NY, USA,
  2018. Association for Computing Machinery.
\newblock ISBN 9781450356343.
\newblock \doi{10.1145/3180445.3180449}.
\newblock URL \url{https://doi.org/10.1145/3180445.3180449}.

\bibitem[Tang et~al.(2014)Tang, Sethumadhavan, and
  Stolfo]{malwareClassificationAnomaly}
Adrian Tang, Simha Sethumadhavan, and Salvatore~J. Stolfo.
\newblock Unsupervised anomaly-based malware detection using hardware features.
\newblock In Angelos Stavrou, Herbert Bos, and Georgios Portokalidis, editors,
  \emph{Research in Attacks, Intrusions and Defenses}, pages 109--129, Cham,
  2014. Springer International Publishing.
\newblock ISBN 978-3-319-11379-1.

\bibitem[Intra2Net(2022)]{PublicBlocklist}
Intra2Net.
\newblock Blacklist monitor, 2022.
\newblock URL \url{https://www.intra2net.com/en/support/antispam/}.

\bibitem[Singh et~al.(2012)Singh, Walenstein, and
  Lakhotia]{malwareConceptDrift}
Anshuman Singh, Andrew Walenstein, and Arun Lakhotia.
\newblock Tracking concept drift in malware families.
\newblock In \emph{Proceedings of the 5th ACM Workshop on Security and
  Artificial Intelligence}, AISec '12, page 81–92, New York, NY, USA, 2012.
  Association for Computing Machinery.
\newblock ISBN 9781450316644.
\newblock \doi{10.1145/2381896.2381910}.
\newblock URL \url{https://doi.org/10.1145/2381896.2381910}.

\bibitem[Pendlebury et~al.(2019)Pendlebury, Pierazzi, Jordaney, Kinder, and
  Cavallaro]{pendlebury2019tesseract}
Feargus Pendlebury, Fabio Pierazzi, Roberto Jordaney, Johannes Kinder, and
  Lorenzo Cavallaro.
\newblock Tesseract: Eliminating experimental bias in malware classification
  across space and time.
\newblock In \emph{28th USENIX Security Symposium (USENIX Security 19)}, pages
  729--746, 2019.

\bibitem[Barbero et~al.(2022)Barbero, Pendlebury, Pierazzi, and
  Cavallaro]{barbero2022transcendent}
Federico Barbero, Feargus Pendlebury, Fabio Pierazzi, and Lorenzo Cavallaro.
\newblock Transcending transcend: Revisiting malware classification in the
  presence of concept drift.
\newblock In \emph{{IEEE} Symposium on Security and Privacy}, 2022.

\bibitem[Fedorchuk and Lamiroy(2017)]{fedorchukclassifier}
Maksym Fedorchuk and Bart Lamiroy.
\newblock Binary classifier evaluation without ground truth.
\newblock In \emph{2017 Ninth International Conference on Advances in Pattern
  Recognition (ICAPR)}, 2017.

\bibitem[Ratner et~al.(2017{\natexlab{a}})Ratner, Sa, Wu, Selsam, and
  Ré]{ratner2017data}
Alexander Ratner, Christopher~De Sa, Sen Wu, Daniel Selsam, and Christopher
  Ré.
\newblock Data programming: Creating large training sets, quickly,
  2017{\natexlab{a}}.

\bibitem[Shiravi et~al.(2012)Shiravi, Shiravi, Tavallaee, and
  Ghorbani]{shiravi2012toward}
Ali Shiravi, Hadi Shiravi, Mahbod Tavallaee, and Ali~A Ghorbani.
\newblock Toward developing a systematic approach to generate benchmark
  datasets for intrusion detection.
\newblock \emph{computers \& security}, 31\penalty0 (3):\penalty0 357--374,
  2012.

\bibitem[Anderson and Roth(2018)]{anderson2018ember}
Hyrum~S Anderson and Phil Roth.
\newblock Ember: an open dataset for training static pe malware machine
  learning models.
\newblock \emph{arXiv preprint arXiv:1804.04637}, 2018.

\bibitem[Arp et~al.(2014)Arp, Spreitzenbarth, Hubner, Gascon, Rieck, and
  Siemens]{arp2014drebin}
Daniel Arp, Michael Spreitzenbarth, Malte Hubner, Hugo Gascon, Konrad Rieck,
  and CERT Siemens.
\newblock Drebin: Effective and explainable detection of android malware in
  your pocket.
\newblock In \emph{Ndss}, volume~14, pages 23--26, 2014.

\bibitem[Marcus and Parameswaran(2015)]{crowdsourcing-book}
Adam Marcus and Aditya Parameswaran.
\newblock \emph{Crowdsourced {D}ata {M}anagement: {I}ndustry and {A}cademic
  {P}erspectives ({B}ook)}.
\newblock Foundations and Trends® in Databases, December 2015.

\bibitem[Crowston(2012)]{mTurk}
Kevin Crowston.
\newblock Amazon mechanical turk: A research tool for organizations and
  information systems scholars.
\newblock In Anol Bhattacherjee and Brian Fitzgerald, editors, \emph{Shaping
  the Future of ICT Research. Methods and Approaches}, pages 210--221, Berlin,
  Heidelberg, 2012. Springer Berlin Heidelberg.
\newblock ISBN 978-3-642-35142-6.

\bibitem[Li et~al.(2019)Li, Dunn, Pearce, McCoy, Voelker, and
  Savage]{li2019reading}
Vector~Guo Li, Matthew Dunn, Paul Pearce, Damon McCoy, Geoffrey~M Voelker, and
  Stefan Savage.
\newblock Reading the tea leaves: A comparative analysis of threat
  intelligence.
\newblock In \emph{28th USENIX Security Symposium (USENIX Security 19)}, pages
  851--867, 2019.

\bibitem[Ratner et~al.(2019)Ratner, Bach, Varma, and R{\'e}]{ratner2019weak}
Alex Ratner, Stephen Bach, Paroma Varma, and Chris R{\'e}.
\newblock Weak supervision: the new programming paradigm for machine learning.
\newblock \emph{Hazy Research. Available via https://dawn. cs. stanford.
  edu//2017/07/16/weak-supervision/. Accessed}, pages 05--09, 2019.

\bibitem[Ratner et~al.(2017{\natexlab{b}})Ratner, Bach, Ehrenberg, Fries, Wu,
  and Ré]{Snorkel2017}
Alexander Ratner, Stephen~H. Bach, Henry Ehrenberg, Jason Fries, Sen Wu, and
  Christopher Ré.
\newblock Snorkel.
\newblock \emph{Proceedings of the VLDB Endowment}, 11\penalty0 (3):\penalty0
  269–282, Nov 2017{\natexlab{b}}.
\newblock ISSN 2150-8097.
\newblock \doi{10.14778/3157794.3157797}.
\newblock URL \url{http://dx.doi.org/10.14778/3157794.3157797}.

\bibitem[sno()]{snorkelcasestudies}
Snorkel case studies.
\newblock https://snorkel.ai/case-studies/.

\bibitem[Tully et~al.(2019)Tully, Haigh, Gibble, and Sikorski]{tully2019sifter}
Philip Tully, Matthew Haigh, Jay Gibble, and Michael Sikorski.
\newblock Learning to rank relevant malware strings using weak supervision.
\newblock \emph{CAMLIS}, 2019.

\bibitem[Deng and Zheng(2021)]{Deng2021CVPR}
Weijian Deng and Liang Zheng.
\newblock Are labels always necessary for classifier accuracy evaluation?
\newblock In \emph{Proceedings of the IEEE/CVF Conference on Computer Vision
  and Pattern Recognition (CVPR)}, pages 15069--15078, June 2021.

\bibitem[Novák et~al.(2019)Novák, Mírovský, Rysová, and
  Rysová]{czechTranslation}
Michal Novák, Jiří Mírovský, Kateřina Rysová, and Magdaléna Rysová.
\newblock \emph{Exploiting Large Unlabeled Data in Automatic Evaluation of
  Coherence in Czech}, pages 197--210.
\newblock 08 2019.
\newblock ISBN 978-3-030-27946-2.
\newblock \doi{10.1007/978-3-030-27947-9_17}.

\bibitem[Dawid and Skene(1979)]{DawidSkene}
A.~P. Dawid and A.~M. Skene.
\newblock Maximum likelihood estimation of observer error-rates using the em
  algorithm.
\newblock \emph{Journal of the Royal Statistical Society. Series C (Applied
  Statistics)}, 28\penalty0 (1):\penalty0 20--28, 1979.
\newblock ISSN 00359254, 14679876.
\newblock URL \url{http://www.jstor.org/stable/2346806}.

\bibitem[Welch(1947)]{welchTest}
B.~L. Welch.
\newblock {The generalization of "Student's" Problem when several different
  populations variances are involved}.
\newblock \emph{Biometrika}, 34\penalty0 (1-2):\penalty0 28--35, 01 1947.
\newblock ISSN 0006-3444.
\newblock \doi{10.1093/biomet/34.1-2.28}.
\newblock URL \url{https://doi.org/10.1093/biomet/34.1-2.28}.

\bibitem[Erdemir et~al.(2021)Erdemir, Bickford, Melis, and
  Aydore]{erdemir2021adversarial}
Ecenaz Erdemir, Jeffrey Bickford, Luca Melis, and Sergul Aydore.
\newblock Adversarial robustness with non-uniform perturbations.
\newblock In M.~Ranzato, A.~Beygelzimer, Y.~Dauphin, P.S. Liang, and J.~Wortman
  Vaughan, editors, \emph{Advances in Neural Information Processing Systems},
  volume~34, pages 19147--19159. Curran Associates, Inc., 2021.

\bibitem[Hucka(2018)]{Hucka2018}
Michael Hucka.
\newblock Nostril: A nonsense string evaluator written in python.
\newblock \emph{Journal of Open Source Software}, 3\penalty0 (25):\penalty0
  596, 2018.
\newblock \doi{10.21105/joss.00596}.
\newblock URL \url{https://doi.org/10.21105/joss.00596}.

\bibitem[Corporation(2022)]{Msdn}
Microsoft Corporation.
\newblock Programming reference for the win32 api, 2022.
\newblock URL \url{https://docs.microsoft.com/en-us/windows/win32/api/}.

\bibitem[M~Sikorski(2012)]{praticalMalware}
A.~Honig M~Sikorski.
\newblock \emph{Practical Malware Analysis}.
\newblock William Pollock, 2012.

\end{thebibliography}

\appendix

\newpage
\section{Supplementary methods for simulated data experiments}
\label{app:simulations}
Here, we outline the generative process used in Section \ref{sec:simulations} to define qualitative conditions under which Firenze can identify the better model with its region-based hypothesis tests. As a reminder, the parameters for our simulated environment are the positive class prevalence $\pi$, model performances on ground-truth and training labels $P_\text{true}^R$, $P_\text{train}^R$,$P_\text{true}^T$, and $P_\text{train}^T$, the number of samples $N$, the region size $K$, and the accuracies and coverages of the marker $\alpha$ and $\beta$ {\it resp.} and the training labels $\bar{\alpha}$ and $\bar{\beta}$ {\it resp.}. Their default values, unless specified otherwise, are $P_\text{train}^T=0.97$, $P_\text{train}^R=0.98$, $P_\text{true}^T=0.95$, $P_\text{true}^R=0.90$, $\pi=0.5$, $\overline{\alpha}=0.95$, $\overline{\beta}=0.10$, $K=10000$, and $N=1000000$. 

\subsection{Label and weak signal generation}
Let $y_\text{true}\in\{-1,1\}$ be the unobserved ground-truth label for a sample $s$, generated as a Bernoulli random variable with probability/bias $\pi$. Let $m \in \{-1,0,1\}$ be the weak label assigned to this sample by a marker. The marker provides a noisy observation of this ground-truth label generated as two more Bernoulli random variables\todo{{\it DONE}: do we need coin analogy here? Can we speak in terms of Bernoulli random variables?}. The first determines whether the marker yields a label; with probability/bias $\beta$, the marker provides a label, otherwise a null-value. The second determines whether the marker yields the {\it correct} label; with probability/bias $\alpha$, the marker takes the actual label $y_\text{true}$, otherwise it flips it\todo{{\it UNK, this is the generative process, which the proof does not use/need}: we don't consider the possbility of flipping for the proofs in the previous section. This might confuse the reader.}. Parametrized this way, $\pi$ defines the {\it prevalence} of the positive class, $\beta$ defines the {\it coverage} of the marker, and $\alpha$ defines its {\it accuracy}. The resulting data-generating process is given by \todo{{\it YES, TODO}: Should we draw a graphical model for this generative process?}
\begin{flalign} 
	\label{eq:marker-sim}
	m\mid y_\text{true}, a, b &= \left\{\begin{matrix} a\cdot y_\text{true} \\ 0\end{matrix}\right.\quad \begin{matrix}\text{if}\:\: b=1 \\ \text{otherwise}\end{matrix}\\
	b&\sim\textsf{\small Bernoulli}(\beta)\\
	a&\sim\textsf{\small Bernoulli}(\alpha)\\
	y_\text{true}&\sim\textsf{\small Bernoulli}(\pi)
\end{flalign}
We simulate this process for each of the $i=1,...,N$ samples $s_i$ and the single marker $m(s_i)$. For this simulated environment, we simulate a single marker, which emulates the most conservative regime of the Firenze framework. \eat{In the previous section, we prove how additional markers---under mild conditions---can only increase the reliability of the marker score.}

Let $y$ be the observed label used for model training. We can simulate the same process for this label, subject to its own coverage $\bar{\beta}$ and accuracy $\bar{\alpha}$. By design, the accuracy of these labels is much higher than that of any markers, $\bar{\alpha}\gg \alpha_j$, but still subject to noise and discrepancies from ground-truth labels:
\begin{flalign}
	y\mid y_\text{true} &= \left\{\begin{matrix} \bar{a}\cdot y_\text{true} \\ 0\end{matrix}\right.\quad \begin{matrix}\text{if}\:\: \bar{b}=1 \\ \text{otherwise}\end{matrix}\\
	\bar{b}&\sim\textsf{\small Bernoulli}(\bar{\beta})\\
	\bar{a}&\sim\textsf{\small Bernoulli}(\bar{\alpha})
\end{flalign}
Neither training labels nor model training are part of Firenze itself; they {\it are} part of our simulation as a means to generate model scores with fully specified performances in the next subsection. These scores become the ``input'' to Firenze.



\subsection{Model score generation} \todo{Bhavna and Mike to go over this section together. I changed the probability ranges to (0,0.49) and (0.51,1) for the model scores to clarify because seems like if we got a 0.5, then it would be an undetermined o/p. We did something similar in the code as well. Let's discuss to ensure we're doing/writing the right thing.}
Let $p\in (0, 1)$ be the model score of a sample $s$, and let $\hat y=\textsf{\small sign}(p-0.5)$ be a decision function that yields a class estimate from that score. This estimate has an observed performance with respect to the training [feed] labels, and an unobserved performance with respect to the ground-truth labels. 

We emulate the training process by generating scores as uniform random variables, with model performance enforced by a Bernoulli random variable with bias/probability $P$. The uniform variable generates noise without affecting the class estimate, drawn between $(0, 0.49)$ if $y=-1$ and $(0.51, 1)$ if $y=1$; for samples without training labels ($y=0$), we use $y_\text{true}$ in its place. The Bernoulli variable determines whether the class estimate is {\it correct}; with probability/bias $P$, the sample remains consistent with the ground-truth or training label, otherwise it flips to the other half-interval. Parametrized this way, $P$ defines the performance of the ``trained'' models; for models $R$ and $T$, across samples {\it with} a training label, we have performances $P_\text{train}^R$ and $P_\text{train}^T$, and for samples {\it without} a training label, we have $P_\text{true}^R$ and $P_\text{true}^T$. The resulting data-generating process---identically for models $R$ and $T$---is given by
\begin{flalign}
	p\mid f, c &= \left\{\begin{matrix} f \\ 1-f\end{matrix}\right.\quad \begin{matrix}\text{if}\:\: c=1 \\ \text{otherwise}\end{matrix}\\
	f\mid y, y_\text{true} &\sim \left\{\begin{matrix} \textsf{\small Uniform}(0.5, 1) \\ \textsf{\small Uniform}(0, 0.5)\end{matrix}\right.\quad \begin{matrix}\text{if}\:\: y=1\:\:\text{or}\:\:y_\text{true}=1\wedge y=0 \\ \text{otherwise}\end{matrix}\\
	c\mid \bar{b}&\sim\left\{\begin{matrix} \textsf{\small Bernoulli}(P_\text{train}) \\ \textsf{\small Bernoulli}(P_\text{true})\end{matrix}\right.\quad \begin{matrix}\text{if}\:\: \bar{b}=1 \\ \text{otherwise}\end{matrix}
\end{flalign}

\section{Supplementary materials for malware detection case study}
\label{app:ember}
\subsection{Marker design rationale for malware detection}
\label{app:ember:markerdetails}
EMBER includes the following groups of raw data describing PE files-- general properties, header information, import functions, export functions section information, byte histogram, byte entropy, and string information. In keeping with our requirement to not use signals which are used for training as markers, here we [artificially] split the available data as follows. We used the general information, sectional header and imports information to design the markers, while the remaining features were used to train the models. This split was necessary; we restrict ourselves to the data available in EMBER for all parts of the experiment to ensure that the study is replicable. The following markers were designed with these fields: 
\begin{itemize}[noitemsep,nolistsep]
\item {\bf Suspicious Section Properties}: In a binary, if we have more than one executable section or we have any sections that are writable and executable, then the file is likely bad. In benign files, it is expected that only the .text section will hold code and be executable. Deviation from this rule of thumb warrants suspicion. And if a file has sections that are writable and executable then that can indicate the presence of self modifying code, which is (malware-like behavior). Thus if a file has more than one section that is Readable/Executable, or any sections that are Writeable/Executable then it is likely malicious.
\item {\bf Unusual Number of Imported Funtions}: Most binaries import multiple libraries and functions. A very low number of imports can indicate packing or some other type of obfuscation. There are exceptions ofcourse; an important one being managed code (written in .Net) where mscoree.dll is often the only import. Looking at a random sampling of benign files we observed that most have on average more than 100 imported functions. Whereas, looking at samples of binaries packed with UPX (a common packing utility used frequently by malware to thwart static analysis and signature matching) we see 5-25 imports. Thus if a file has less than 25 import functions, then we deem it as likely malware.
\item {\bf Nonsensical Section Names}: Windows binaries usually contain multiple sections. Most commonly, one or more of the following are present. .text, .rdata, .data, .rsrc, .reloc. Less frequently, but still prevalent are others like .idata, .edata, .pdata or CODE. On the other hand, there are section names that warrant suspicion. For example .UPX (and variants thereof) are added by UPX. While not all files packed with UPX are malware, from past experience, we know that it is heavily used by malware authors. Additionally, files with nonsensical section names are also likely to be malware and not legitimate software. To detect whether a section name is “nonsensical” we use nostril [reference: https://joss.theoj.org/papers/10.21105/joss.00596]. If we see known suspicious section names, or nonsensical section names in a file, we deem it likely malware.
\item {\bf Import of suspicious functions}: There are certain functions and libraries that are used by binaries to implement functionality that is likely to be associated with malware. Thus presence of these functions in the imports of a binary makes it suspicious. We use the presence of such functions as a test of suspiciousness in this marker.  For example, process injection which is commonly used by malware to elevate privileges or access resources belonging to another process exhibits some peculiar function call patterns. The malware might call a series of Process32First/Next and Thread32First/Next to identify the process or thread it wants to inject in and then call VirtualAllocEx to allocate memory in the remote process. Thus the presence of these functions in the imports section of a binary makes it suspicious. Ofcourse, there are behavior that a malware might exhibit that will also be common amongst benign files. CreateFile is such a function that is used broadly by malware and benignware. Expertise and experience is required to design this marker. In Appendix A we share a table of the functions we used in this markers and how they are used by malware. 
\item {\bf Signed}: A signed PE file indicates that the file is from a trusted source and is likely benign. While there are samples of signed malware (for example, expired certificates stolen during NVIDIA’s compromise by the Lapsus\$ group were used to sign malware.
\end{itemize}

\subsection{Table of Functions Used in the Suspicious Imports Marker}
\label{app:ember:importfunctions}
This table describes the functions used to define the Suspicious Imports Marker. The descriptions were obtained from MSDN \citep{Msdn}. Existing resources like \citep{praticalMalware} can be used to obtain such a list.
\begin{table}[H]
	\centering
	\caption{Functions Used in the Suspicious Imports Marker}
	\label{tab:ImportFunctions}
			\begin{tabular}{l|p{0.6\linewidth}|p{0.3\linewidth}} 
					\textbf{Function} & \textbf{Description}  & \textbf{Tactic or Type of Malware Associated} \\
					\hline
					createprocessasuser	&	Creates a new process and its primary thread. The new process runs in the security context of the user represented by the specified token.	&	Injection	\\
					createservice	&	After openscmanager, this is used to create the service which will run the malware functionality at startup	&	Persistence	\\
					cryptbinarytostring	&	Converts an array of bytes into a formatted string	&	Ransomware	\\
					cryptcreatehash	&	Initiates the hashing of a stream of data	&	Ransomware	\\
					cryptdestroyhash	&	Destroys the hash object	&	Ransomware	\\
					cryptgethashparam	&	Get the hashed value after applying an algorithm	&	Ransomware	\\
					crypthashdata	&	The CryptHashData function adds data to a specified hash object	&	Ransomware	\\
					encryptfile	&	Encrypt a file or directory	&	Ransomware	\\
					getadaptersinfo	&	Used to obtain information about network adapters. Can be recon, or check for anti-vm functionality	&	Anti VM Functionality	\\
					getforegroundwindow	&	Returns Handle to the window that is in the foreground. Used by keyloggers to determine which window the user is entering key strokes into	&	Keylogger	\\
					internetopen	&	Initializes internet access functions from WinINet	&	C2 functionality	\\
					mapvirtualkey	&	Translates virtual keycode into a character value	&	Keylogger	\\
					process32first	&	Used to enumerate processes by malware prior to injection	&	Process Injection	\\
					process32next	&	Used to enumerate processes by malware prior to injection	&	Process Injection	\\
					regopenkey	&	Opens a handle to read or edit a registry key which is a common persistence mechanism	&	Persistence	\\
					regsavekey	&	Saves the specified key and all of its subkeys and values to a new file, in the standard format.	&	Persistence	\\
					setprop	&	Used by malware to register a property and wait for its invocation to execute malicious commands.	&	Process Injection	\\
					thread32first	&	Used to enumerate threads prior to injection	&	Process Injection	\\
					thread32next	&	Used to enumerate threads prior to injection	&	Process Injection	\\
					urldownloadtofile	&	Download a file from a webserver	&	Downloader	\\
					virtualallocex	&	Allocates memory in a remote process	&	Process Injection	\\
					virtualprotectex	&	Changes the protection on a memory region to make it executable	&	Process Injection	\\
					winexec	&	Execute a new program	&	Downloader	\\
				\end{tabular}
\end{table}

\section{Supplementary materials for domain name reputation case study}
\label{app:mithra}
\subsection{Marker design rationale for domain reputation}
\label{app:mithra:markers}

Recall the seven markers introduced in the main text:
		\begin{longtable}{l|l}
			Abused Domain &	If the domain is associated with curated abused domain list, then 1, else 0 \\
			Sinkholed Domain &	If the domain is associated with curated sinkhole IP address list, then 1, else 0 \\
			Honeypot Domain &	If the domain appears in in-house honeypot logs, then 1, else 0 \\
			Domain Popularity &	If the domain is considered popular based on query counts, then -1, else 0 \\
			Number of  IPs	& If the domain maps to more than 50 unique IP addresses, then -1, else 0 \\
			Number of TTLs &	If the domain appears with more than 500 TTLs, then -1, else 0 \\
			Known Future Label	& If the domain is labeled malicious in the future labels, then 1, \\ 
			& if it is labeled benign, then -1, else 0
		\end{longtable} 

Based on past manual analysis, one interesting signal of maliciousness we found is the association with abused top-level domains (TLDs) and effective second level domain names (e2LD, the smallest unit of a domain name that can be registered by Internet users). Owners of these TLDs and e2LDs allow actors to register domain names for free or minimal cost. Though being related to an abused TLD is a good signal of suspiciousness, legitimate domains also exist within these name spaces and not all domains associated with abused TLDs and e2LDs should be considered malicious. Thus, while this principle would be bad for labeling domains, it is a great marker. Another example we use is association of a domain with a manually curated list of sinkhole IP addresses. 
Along with malicious markers, we also utilize benign markers. For example, we expect that highly popular domains based on query counts will more likely be benign compared to malicious domains. Popularity itself is not a guarantee of benignity, but a decent signal, and therefore is another good marker. Further examples of benign markers include those domains that resolve to a very high number of IP addresses with multiple different TTLs (Time To Live) — based on our observations, these tend to be associated with Content Delivery Networks (or CDNs) and heavily skew benign. While designing these markers, we also looked at domains that may resolve to a high number of IP addresses due to fast-flux behavior (and therefore are likely malicious), but we observed that the number of unique IPs observed for those domains over a day were far lower than we see for domains associated with CDNs. The thresholds for these markers effectively separate these types of activity. Thus we can see that designing markers is a combination of domain expertise verified by data, art and science. Marker functions will return 1 when the marker considers a domain to be likely malicious and -1 when the marker considers a domain to be likely benign. 0 indicates the marker did not vote. It is important to note that in these experiments, the markers are not used as label sources or features in training. For example, though we have a manually curated list of known sinkhole IP addresses and abused TLDs, these are not used for training as manually maintaining a fully accurate list over time is challenging and we do not want this model to overfit on those types of domain names. The Known Future Label marker is based on what the labels say about a domain one week after the training time. Usually in the security domain we see that we don't have perfect signal about new entities, but within a few days, labels get updated— whether through manual investigations, correlations, or gathering external intelligence. Since these are "Future Labels", they can't be used for training, but are excellent for evaluation. 

\subsection{Fine-grained investigations of Results based on individual markers}
\label{app:mithra:indivmarkers}

We explore two tables of detailed test results here to illustrate how individual markers can be used to explain and dive deeper into the results seen in the summary view provided by combined marker scores. For completeness, we provide all six such tables for $2\times 3$ cases of $K$ = 10k and 100k as well as the three region-based hypothesis tests.

Looking at the detail view of the Top$K$ test over $10k$ region in table \ref{tab:top10k}, we see that the average marker score for the malicious markers (Abused domains and Sinkholed domains) is higher for the reference model than the test model. The differences pass the statistical significance test. This shows that the test model is finding fewer likely malicious domains of these types in its K-most-malicious-domains region than the reference model. We observe a similar outcome in the Known Future Labels Marker as well, where the reference mean is \emph{higher} than the test mean, and the difference is statistically significant. This implies that the test model is detecting fewer domains that will be likely labeled malicious in the future than the reference model. Thus we can say that the test model does not accomplish our stated goal of increasing detection value.  On the benign marker side (Domain Popularity, Number of IPs and Number of TTLs), we observe that the Top$K$ regions of both models are not significantly different.  This implies that there are no likely benign domains that are deemed highly malicious by either model. The results from these markers indicate that the real-world FP rate for the detections from both models are likely to be similar, and the test model preserves the low-FP quality of the reference model. With these data points we can show with a high degree of explainability that the test model is not performing better than the reference model at scoring malicious domains.

Looking at the malicious markers (Abused domains and Sinkholed domains) in the detail view of the Bottom$K$ test results over the $10k$ region in table \ref{tab:bot10k} , we see that the while the means for the test model are consistently lower than the reference model, often the averages for both models are close to zero, and not statistically significant. This implies the FN rate from the benign list generated by both models will be similar. For the benign markers in the same test, we see that the test mean is significantly lower than the reference mean for all markers. This indicates that the test model is finding more likely benign domains than the reference model. Thus, we can once again illustrate our summary result that the test model is better at scoring benign domains.


\begin{table}[H]
	\centering
	\caption{Top K test (K=10,000)}
	\label{tab:top10k}
	\resizebox{0.475\textwidth}{!}{%
		\begin{tabular}{l|r|r|r|c} 
			\textbf{Marker} & \textbf{Avg Marker Score}  & \textbf{Avg Marker Score}  & \textbf{p-value} & \textbf{Result} \\
			& \textbf{Reference Model}  & \textbf{Test Model}  &  &  \\
			\hline
			AbusedDomain	&	0.310569	&	0.294771	&	2.07E-02	&	\color{red}F	\\
			SinkholedDomain	&	0.062194	&	0.020898	&	1.26E-47	&	\color{red}F	\\
			HoneypotDomain	&	0	&	0	&	NaN	&	\color{YellowOrange}U	\\
			DomainPopularity	&	0	&	0	&	NaN	&	\color{YellowOrange}U	\\
			NumberIPs	&	0	&	0	&	NaN	&	\color{YellowOrange}U	\\
			NumberTTLs	&	0	&	0	&	NaN	&	\color{YellowOrange}U	\\
			KnownFutureLabel	&	0.272273	&	0.217278	&	6.88E-19	&\color{red}F	\\
			CombinedMarkerScore	&	0.617138	&	0.516348	&	4.79E-46	&	\color{red}F	\\
		\end{tabular}
	}
\end{table}

\begin{table}[H]
	\centering
	\caption{Bottom K Test (K=10,000)}
	\label{tab:bot10k}
	\resizebox{0.475\textwidth}{!}{%
		\begin{tabular}{l|r|r|r|c}
			\textbf{Marker} & \textbf{Avg Marker Score}  & \textbf{Avg Marker Score}  & \textbf{p-value} & \textbf{Result} \\
			& \textbf{Reference Model}  & \textbf{Test Model}  &  &  \\
			\hline
			AbusedDomain	&	0	&	0	&	NaN	&	\color{YellowOrange}U	\\
			SinkholedDomain	&	0	&	0	&	NaN	&	\color{YellowOrange}U	\\
			HoneypotDomain	&	0.0001	&	0	&	0.241959	&	\color{YellowOrange}U	\\
			DomainPopularity	&	-0.1875	&	-0.7806	&	0.00E+00	&	\color{green}S	\\
			NumberIPs	&	-0.2614	&	-0.669	&	0.00E+00	&	\color{green}S	\\
			NumberTTLs	&	-0.0631	&	-0.3632	&	0.00E+00	&	\color{green}S	\\
			KnownFutureLabel	&	-0.4275	&	-0.7642	&	0.00E+00	&	\color{green}S	\\
			CombinedMarkerScore	&	-0.5795	&	-0.9835	&	0.00E+00	&	\color{green}S	\\
		\end{tabular}
	}
\end{table}

\begin{table}[H]
	\centering
	\caption{Up-Movers and Down-Movers Test (K=10,000)}
	\label{tab:movers10k}
	\resizebox{0.475\textwidth}{!}{%
		\begin{tabular}{l|r|r|r|c}
			\textbf{Marker} & \textbf{Avg Marker Score}  & \textbf{Avg Marker Score}  & \textbf{p-value} & \textbf{Result} \\
			& \textbf{Up-Movers}  & \textbf{Down-Movers}  & &  \\
			\hline
			AbusedDomain	&	0	&	0	&	NaN	&	\color{YellowOrange}U	\\
			SinkholedDomain	&	0	&	0	&	NaN	&	\color{YellowOrange}U	\\
			HoneypotDomain	&	0	&	0	&	NaN	&	\color{YellowOrange}U	\\
			DomainPopularity	&	-0.0006	&	-0.0038	&	3.44E-06	&	\color{green}S	\\
			NumberIPs	&	0	&	-0.0003	&	8.90E-02	&	\color{YellowOrange}U	\\
			
			NumberTTLs	&	-0.0006	&	-0.004	&	1.35E-06	&	\color{green}S	\\
			KnownFutureLabel	&	0.0033	&	0.0133	&	4.32E-09	&	\color{red}F	\\
			CombinedMarkerScore	&	0.0026	&	0.0074	&	1.06E-02	&	\color{red}F	\\
			
		\end{tabular}
	}
\end{table}


\begin{table}[H]
	\centering
	\caption{Top K test (K=100,000)}
	\label{tab:top100k}
	\resizebox{0.475\textwidth}{!}{%
		\begin{tabular}{l|r|r|r|c}
			\textbf{Marker} & \textbf{Avg Marker Score}  & \textbf{Avg Marker Score}  & \textbf{p-value} & \textbf{Result} \\
			& \textbf{Reference Model}  & \textbf{Test Model}  &  &  \\
			\hline
			AbusedDomain	&	0.088239	&	0.063599	&	4.39E-95	&	\color{red}F
			\\
			SinkholedDomain	&	0.237948	&	0.125499	&	0.00E+00	&	\color{red}F
			\\
			HoneypotDomain	&	0	&	0	&	NaN	&	\color{YellowOrange}U
			\\
			DomainPopularity	&	-0.00017	&	-0.00014	&	3.45E-01	&	\color{YellowOrange}U
			\\
			NumberIPs	&	0	&	-0.00001	&	2.42E-01	&	\color{YellowOrange}U
			\\
			NumberTTLs	&	-0.00018	&	-0.00014	&	3.11E-01	&	\color{YellowOrange}U
			\\
			KnownFutureLabel	&	0.268607	&	0.235748	&	2.86E-63	&	\color{red}F
			\\
			CombinedMarkerScore	&	0.570214	&	0.405806	&	0.00E+00	&	\color{red}F \\
		\end{tabular}
	}
\end{table}

\begin{table}[H]
	\centering
	\caption{Bottom K Test (K=100,000)}
	\label{tab:bot100k}
	\resizebox{0.475\textwidth}{!}{%
		\begin{tabular}{l|r|r|r|c}
			\textbf{Marker} & \textbf{Avg Marker Score}  & \textbf{Avg Marker Score}  & \textbf{p-value} & \textbf{Result} \\
			& \textbf{Reference Model}  & \textbf{Test Model}  &  &  \\
			\hline
			AbusedDomain	&	0	&	0	&	NaN	&	\color{YellowOrange}U
			\\
			SinkholedDomain	&	0	&	0.00001	&	2.42E-01	&	\color{YellowOrange}U
			\\
			HoneypotDomain	&	0.00003	&	0.00002	&	3.61E-01	&	\color{YellowOrange}U
			\\
			DomainPopularity	&	-0.23787	&	-0.49372	&	0.00E+00	&	\color{green}S
			\\
			NumberIPs	&	-0.16872	&	-0.19044	&	6.84E-36	&	\color{green}S
			\\
			NumberTTLs	&	-0.08835	&	-0.22822	&	0.00E+00	&	\color{green}S
			\\
			KnownFutureLabel	&	-0.43925	&	-0.46809	&	1.46E-37	&	\color{green}S
			\\
			CombinedMarkerScore	&	-0.54655	&	-0.67804	&	0.00E+00	&	\color{green}S\\
			
		\end{tabular}
	}
\end{table}

\begin{table}[H]
	\centering
	\caption{Up-Movers and Down-Movers Test (K=100,000)}
	\label{tab:movers100k}
	\resizebox{0.475\textwidth}{!}{%
		\begin{tabular}{l|r|r|r|c}
			\textbf{Marker} & \textbf{Avg Marker Score}  & \textbf{Avg Marker Score}  & \textbf{p-value} & \textbf{Result} \\
			& \textbf{Up-Movers}  & \textbf{Down-Movers}  & &  \\
			\hline
			AbusedDomain	&	0	&	0	&	NaN	&	\color{YellowOrange}U
			\\
			SinkholedDomain	&	0	&	0.00002	&	1.47E-01	&	\color{YellowOrange}U
			\\
			HoneypotDomain	&	0	&	0	&	NaN	&	\color{YellowOrange}U
			\\
			DomainPopularity	&	-0.00008	&	-0.00059	&	1.47E-09	&\color{green}S
			\\
			NumberIPs	&	-0.00003	&	-0.00005	&	3.11E-01	&	\color{YellowOrange}U
			\\
			NumberTTLs	&	-0.00006	&	-0.00091	&	2.62E-17	&	\color{green}S
			\\
			KnownFutureLabel	&	0.00048	&	0.00138	&	2.88E-04	&	\color{red}F \\
			CombinedMarkerScore	&	0.00036	&	0.00016	&	2.96E-01	&	\color{YellowOrange}U\\
		\end{tabular}
	}
\end{table}

\end{document}